\documentclass[10pt]{article}
\usepackage{graphicx} 

\begin{document}

\parindent=1.0cm 
\begin{center}

{\large \bf Analytical special solutions of the Bohr Hamiltonian}

\bigskip
{\large 

{\underline{Dennis Bonatsos}, D. Lenis, D. Petrellis}

{Institute of Nuclear Physics, N.C.S.R.
``Demokritos'', GR-15310 Aghia Paraskevi, Attiki, Greece}

{P. A. Terziev} 

{Institute for Nuclear Research and Nuclear Energy, Bulgarian
Academy of Sciences, 72 Tzarigrad Road, BG-1784 Sofia, Bulgaria}

{I. Yigitoglu} 

{Hasan Ali Yucel Faculty of Education, Istanbul University, TR-34470 Beyazit, 
Istanbul, Turkey} 

} 

\end{center}

\centerline{Abstract}
 
{\it The following special solutions of the Bohr Hamiltonian are briefly 
described: 1) Z(5) (approximately separable solution in five dimensions 
with $\gamma \simeq 30^{\rm o}$), 2) Z(4) (exactly separable $\gamma$-rigid 
solution in four dimensions with $\gamma=30^{\rm o}$), 3) X(3) (exactly 
separable $\gamma$-rigid solution in three dimensions with $\gamma=0$). 
The analytical solutions obtained using Davidson potentials in the E(5), 
X(5), Z(5), and Z(4) frameworks are also mentioned.
}

\bigskip

Critical point symmetries \cite{IacE5,IacX5}, describing nuclei at 
points of shape phase transitions between different limiting symmetries,
have recently attracted considerable attention, since they lead to 
parameter independent (up to overall scale factors) predictions which 
are found to be in good agreement with experiment 
\cite{CZ1,Clark1,Pd102,CZ2,Clark2}. The E(5) 
critical point symmetry \cite{IacE5} is supposed 
to correspond to the transition from vibrational [U(5)] to $\gamma$-unstable 
[O(6)] nuclei, while the X(5) critical point symmetry \cite{IacX5}
is assumed to describe 
the transition from vibrational [U(5)] to prolate axially 
symmetric [SU(3)] nuclei. Both symmetries are obtained as special solutions 
of the Bohr Hamiltonian \cite{Bohr}. In the E(5) case \cite{IacE5} 
the potential is supposed to 
depend only on the collective variable $\beta$ and not on $\gamma$. 
Then exact separation of variables is achieved and the equation containing 
$\beta$ can be solved exactly \cite{IacE5,Wilets} for an infinite square well 
potential in $\beta$, the eigenfunctions being Bessel 
functions of the first kind, while the equation containing the angles has been 
solved a long time ago by B\`es \cite{Bes}. 
In the X(5) case \cite{IacX5}
the potential is supposed to be of the form $u(\beta)+u(\gamma)$.
Then approximate separation of variables is achieved in the special case
of $\gamma \simeq 0$, the $\beta$-equation with an infinite square 
well potential 
leading to Bessel eigenfunctions, while the $\gamma$-equation with a harmonic 
oscillator potential having a minimum at $\gamma=0$ leads to a two-dimensional
harmonic oscillator with Laguerre eigenfunctions \cite{IacX5}. 
In both cases the full five variables 
of the Bohr Hamiltonian \cite{Bohr} (the collective variables $\beta$ and 
$\gamma$, 
as well as the three Euler angles) are involved. The algebraic structure 
of E(5) is clear, since the Hamiltonian is the second order Casimir operator 
of E(5), which corresponds to the square of the momentum operator in five 
dimensions (see \cite{E5,Z4} and references therein), 
while an SO(5) subalgebra (generated by the angular momentum operators 
in five dimensions) exists. The algebraic structure of X(5) (if any, since 
X(5) is an approximate and not an exact solution) has not been
identified yet.  

It is of interest to identify additional special cases leading to analytical 
solutions of the Bohr Hamiltonian, and to examine their relation to 
critical behaviour of nuclei. 

It has been known for a long time that the Bohr equation gets simplified 
in the special case of $\gamma=30^{\rm o}$ \cite{Brink,MtVNPA}, since two of 
the principal moments
of inertia become equal in this case, guaranteeeing the existence of a good 
quantum number (the projection $\alpha$ of angular momentum on the body-fixed 
$x$ axis), although the nucleus possesses a triaxial shape. In other words, 
the Hamiltonian possesses a symmetry, while the shape of the nucleus 
does not.  By allowing
the potential to be of the form $u(\beta)+u(\gamma)$, and by permitting 
$\gamma$ to vary only around $\gamma\simeq 30^{\rm o}$, approximate 
separation of variables is achieved \cite{Z5}, similar in spirit to the X(5) 
solution. The $\beta$-equation with an infinite square well potential leads 
then to Bessel eigenfunctions, while the $\gamma$-equation with a harmonic 
oscillator potential having a minimum at $\gamma=30^{\rm o}$ takes the 
form of a simple harmonic oscillator equation. The full five variables of the 
Bohr Hamiltonian are involved in this case, while the algebraic structure 
(if any, since the solution is approximate) is yet unknown. This solution 
has been called Z(5) \cite{Z5}. The relevant level scheme is shown 
in Fig. 1(a). 

\begin{figure}[htb]  
\includegraphics[height=95mm]{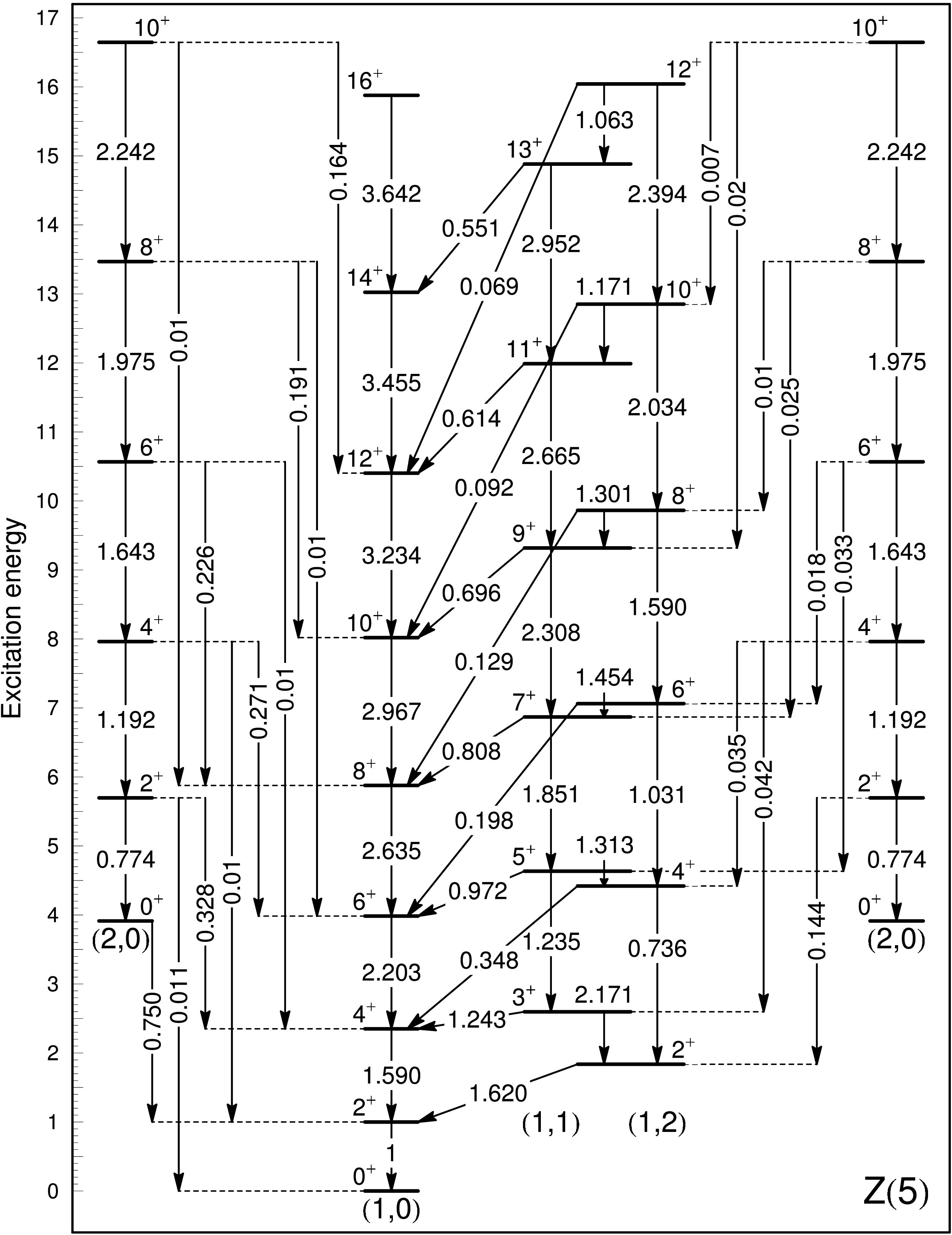}
\includegraphics[height=95mm]{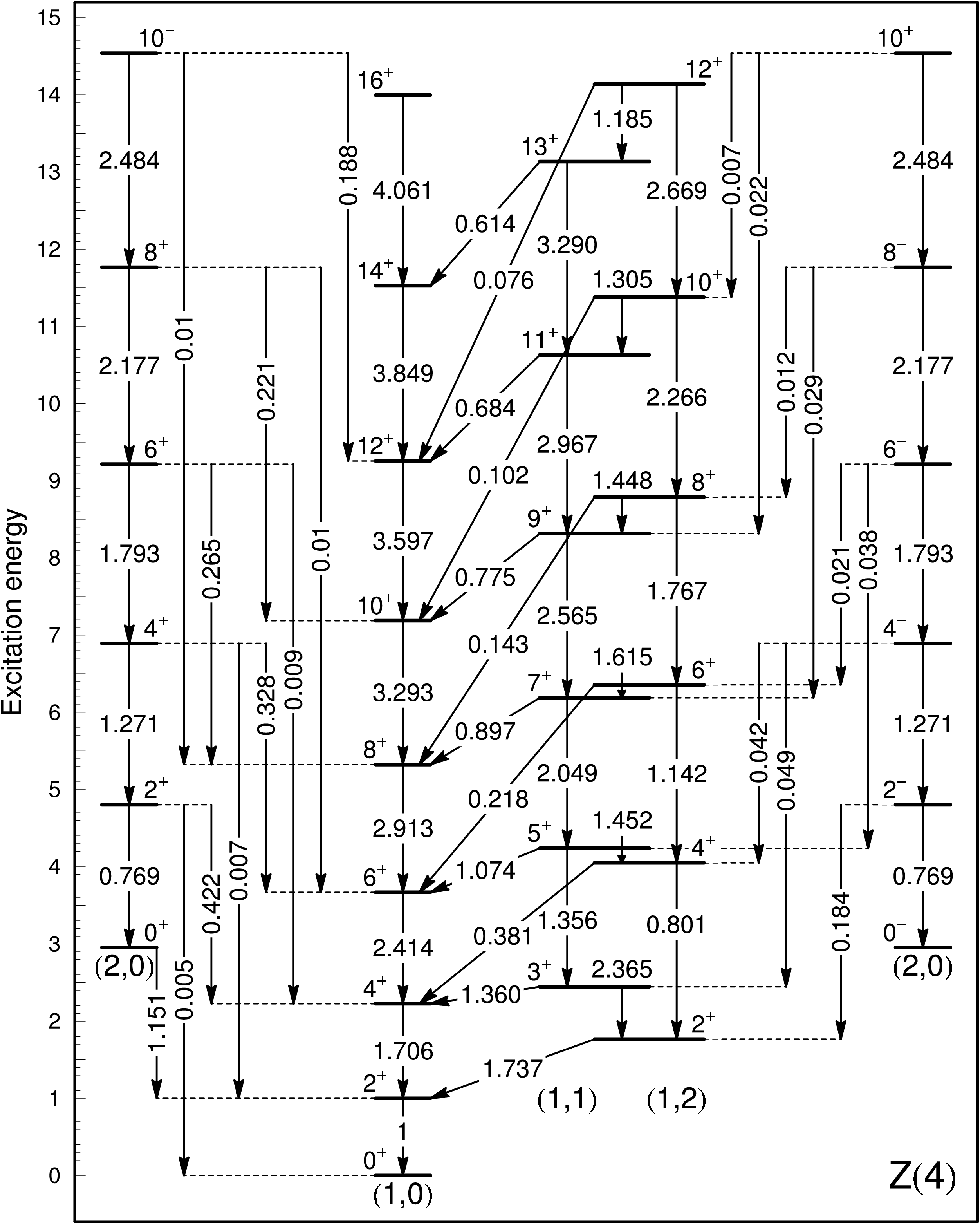}
\caption{(a) 
Intraband and interband B(E2) transition rates in the Z(5) 
model \cite{Z5}, normalized to the B(E2;$2_{1,0}\to 0_{1,0}$) rate. Bands are 
labelled by $(s,n_w)$, their levels being normalized to $2_{1,0}$.  
The (2,0) band is shown both at the left and at 
the right end of the figure for drawing purposes.  
(b) Same for the Z(4) model \cite{Z4}. 
\label{f1}}
\end{figure}

Separation of variables becomes exact by ``freezing'' the $\gamma$ variable 
to the special value of $\gamma=30^{\rm o}$, in the spirit of the 
Davydov and Chaban \cite{Chaban} approach. Then the $\beta$-equation with 
an infinite square well potential leads to Bessel eigenfunctions \cite{Z4}, 
while 
the equation involving the Euler angles and the parameter $\gamma$ (which is
not a variable any more) leads to the solution obtained by Meyer-ter-Vehn 
\cite{MtVNPA}. The  projection 
$\alpha$ of angular momentum on the body-fixed $x$ axis is a good quantum 
number also in this case. Only four 
variables ($\beta$ and the three Euler angles) are involved,
while the full algebraic structure is yet unknown. It has been remarked
\cite{Z4}, however, that the ground state band of this model coincides with 
the ground state band of E(4), the euclidean algebra in four dimensions. 
This solution has been labelled as Z(4) \cite{Z4}.  The relevant level scheme 
is shown in Fig. 1(b), while in Fig. 2(a) the great similarity of 
the ground state band and the $\beta_1$-band of the Z(4) model to the 
corresponding bands of E(5) is demonstrated. The main difference between 
the two models occurs in the $\gamma_1$ band, as seen in Fig. 2(b).
Experimental examples of Z(4) seem to appear in $^{128-132}$Xe, as shown 
in Fig. 3, while experimental manifestations of Z(5) seem to appear 
in $^{192-196}$Pt, as shown in Fig. 4.  

\begin{figure}[htb]  
  \rotatebox{270}{\includegraphics[width=55mm]{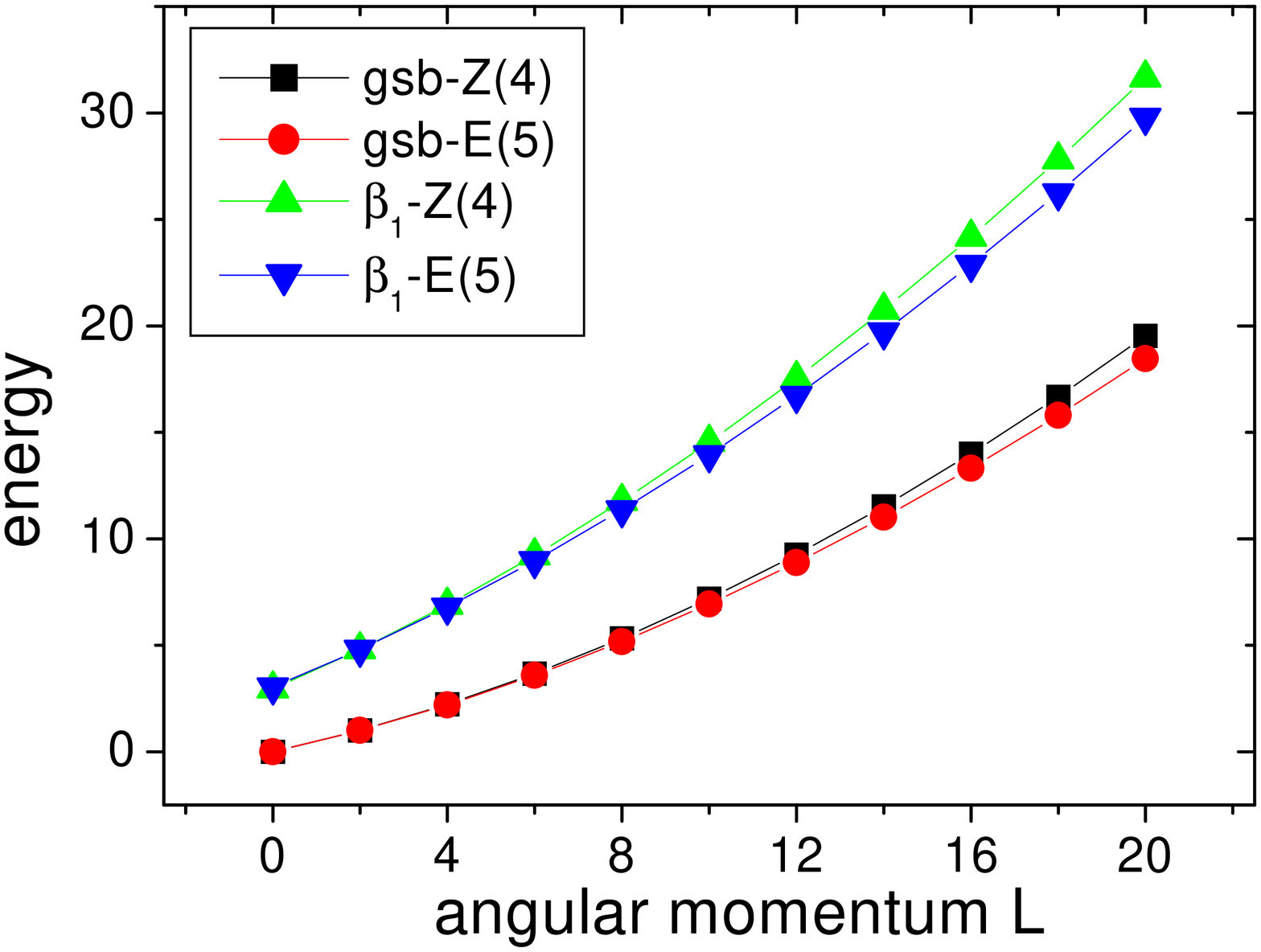}}
\rotatebox{270}{\includegraphics[width=55mm]{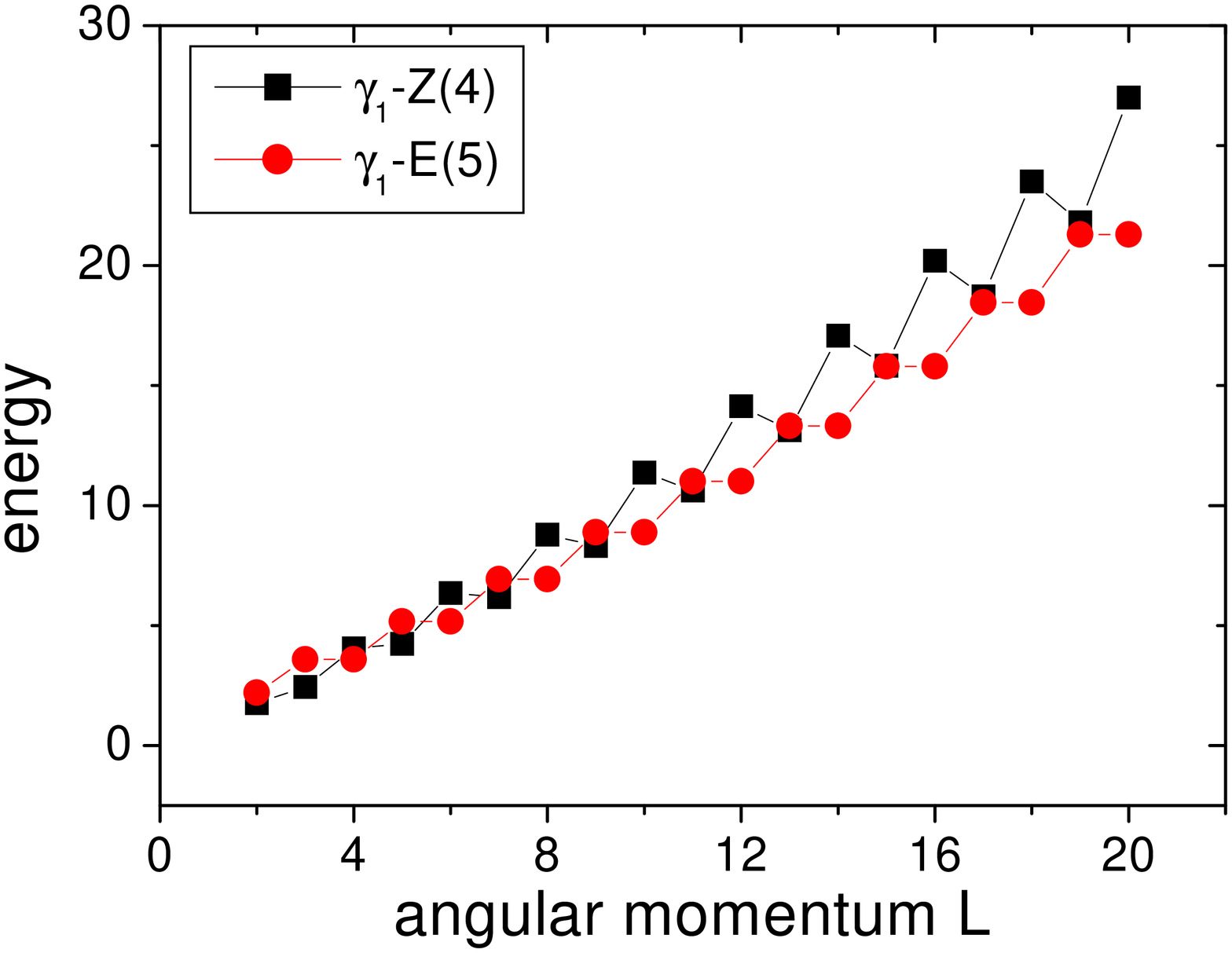}}
\caption{(a) Ground state band [$(s,n_w)=(1,0)$] and first excited band 
[$(s,n_w)=(2,0)$] of Z(4) \cite{Z4} (labeled as $\beta_1$-band)
compared to the corresponding bands of E(5) \cite{IacE5,E5}. In each model 
all levels are normalized to the $2_1^+$ state. 
(b) The lowest``$K=2$ band'' of Z(4) [formed out of the ($s,n_w$) bands 
(1,2) and (1,1), labeled as $\gamma_1$], compared to the corresponding band 
of E(5).
\label{f2}}
\end{figure}

\begin{figure}[htb]   
\includegraphics[width=135mm]{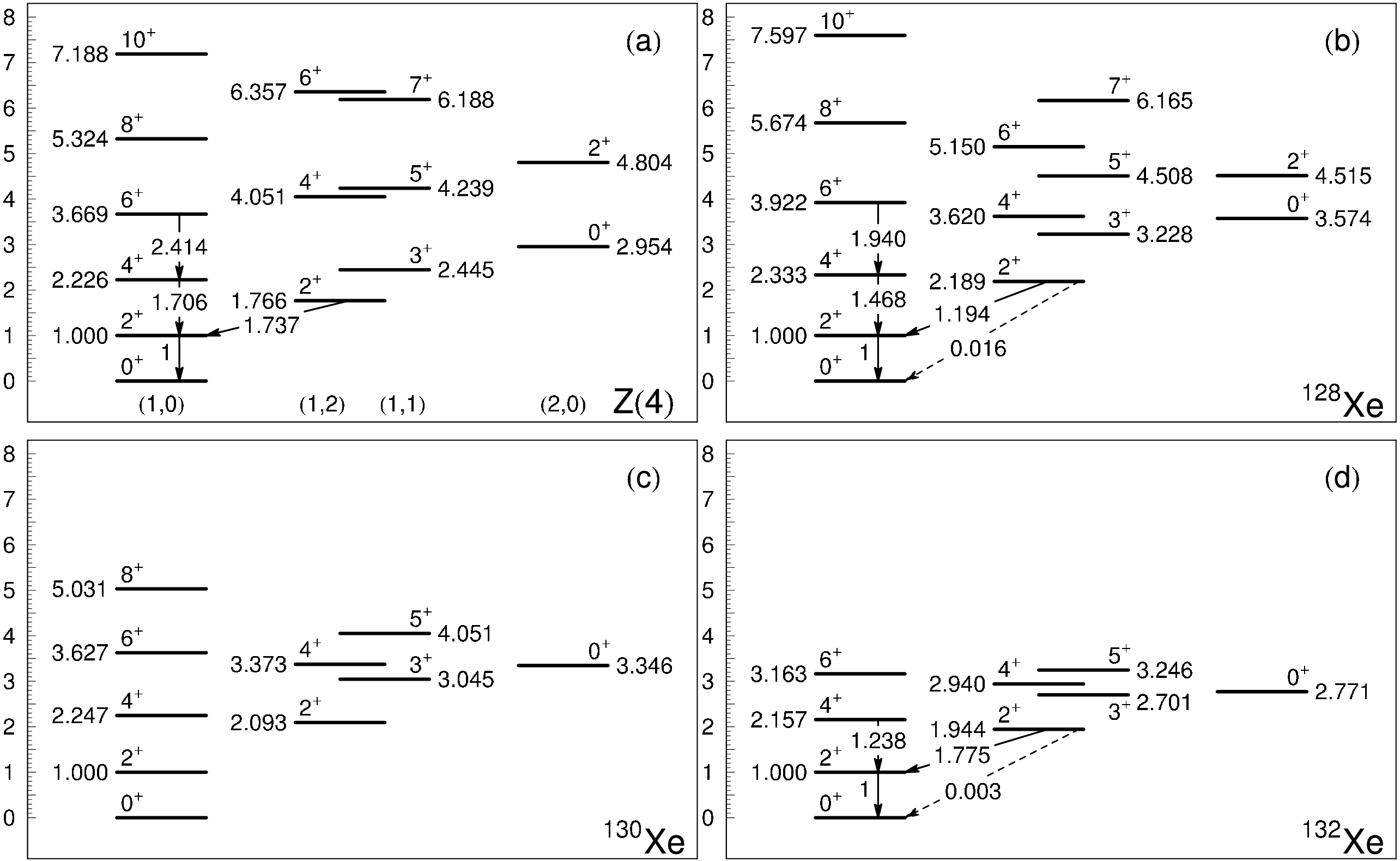}
\caption{Comparison of the Z(4) predictions \cite{Z4} for (normalized) 
energy levels 
and (normalized) B(E2) transition rates (a) to experimental data for 
$^{128}$Xe \cite{Xe128} (b), $^{130}$Xe \cite{Xe130} (c), 
and $^{132}$Xe \cite{Xe132} (d). Bands in (a) are labelled by $(s,n_w)$.  
\label{f3}}
\end{figure}

\begin{figure}[htb]   
\includegraphics[width=135mm]{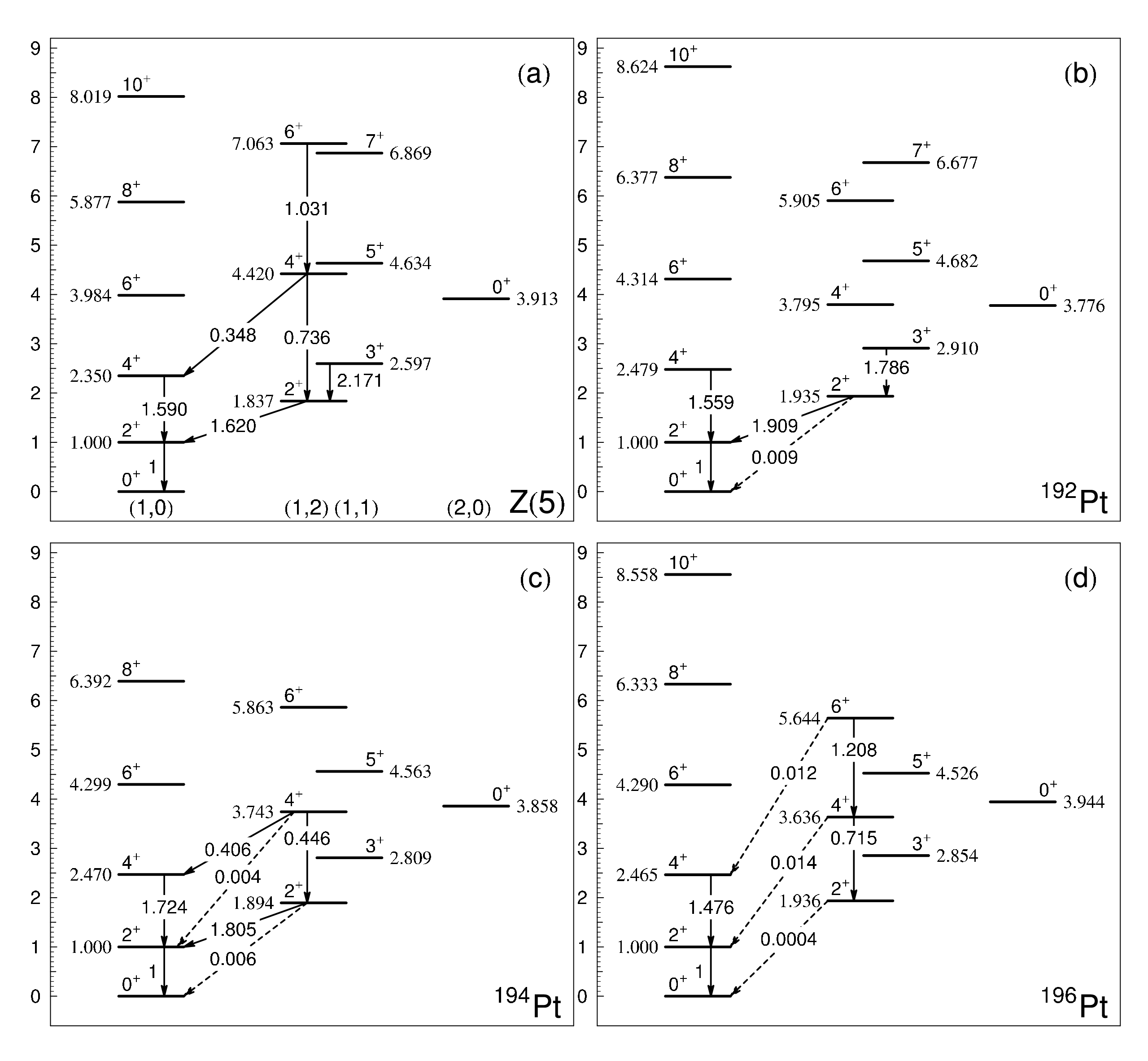}
\caption{Comparison of the Z(5) predictions \cite{Z5} for (normalized) energy 
levels and (normalized) B(E2) transition rates (a) to experimental data for 
$^{192}$Pt \cite{Pt192} (b), $^{194}$Pt \cite{Pt194} (c), 
and $^{196}$Pt \cite{Pt196} (d). Bands in (a) are labelled by $(s,n_w)$.  
\label{f4}}
\end{figure}

The question arises then of what happens in the case one ``freezes'' the 
$\gamma$ variable to the value $\gamma=0$, which corresponds to axially 
symmetric prolate shapes, for which the projection $K$ of angular momentum 
on the body-fixed $z$-axis is a good quantum number. It turns out \cite{X3}
that 
only three degrees of freedom are relevant in this case, since the nucleus 
is axially symmetric, so that two angles suffice for describing its direction 
in space, while the variable $\beta$ describes its shape. Separation 
of variables becomes exact \cite{X3}, the $\beta$ equation with an infinite 
square 
well potential leading to Bessel eigenfunctions, while the equation involving 
the angles leads to the simple spherical harmonics. The algebraic structure 
of this model is yet unknown. This solution has been called X(3) \cite{X3}. 
The level scheme is shown in Fig. 5 and Table 1. 
Experimental examples of X(3) seem to occur in $^{172}$Os and $^{186}$Pt, 
as seen in Fig. 6, in which also X(5) is compared to experimental data 
for the N=90 isotones $^{150}$Nd \cite{150Nd}, $^{152}$Sm \cite{152Sm}, 
$^{154}$Gd \cite{154Gd}, and $^{156}$Dy \cite{156Dy}, which are known 
to be very good examples of the X(5) critical point symmetry 
\cite{CZ2,Kruecken,Tonev,Dewald,CaprioDy}.  
A curious coincidence of the $\beta_1$-bands of these nuclei with the X(3) 
predictions occurs.  

\begin{figure}[htb]   
\includegraphics[height=65mm]{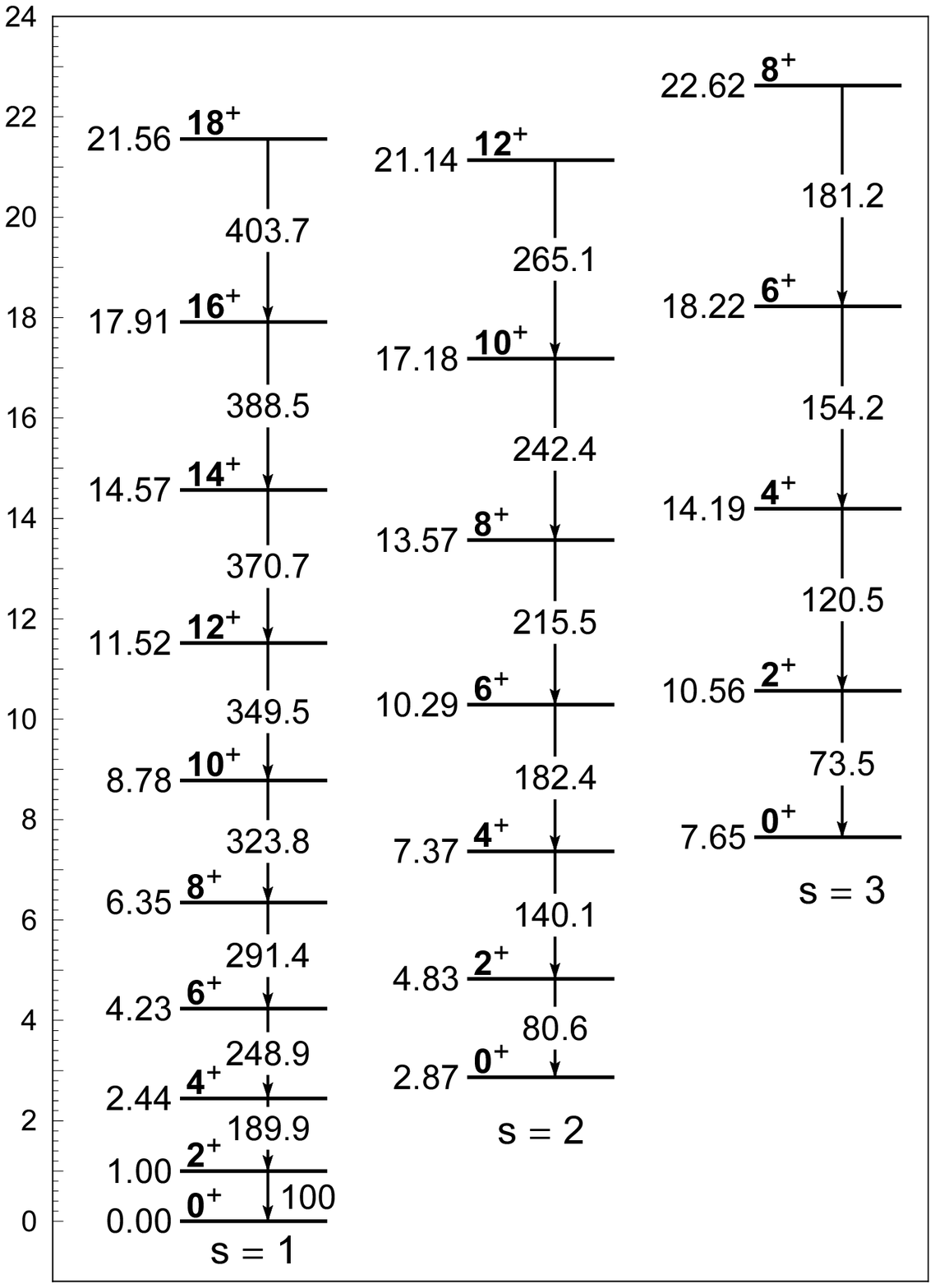}
\caption{Energy levels of the ground state ($s=1$), $\beta_1$ ($s=2$), 
and $\beta_2$ ($s=3$) bands of X(3) \cite{X3}, normalized to the energy 
of the lowest excited state, $2_1^+$, together with 
intraband $B(E2)$ transition rates, normalized to the transition between 
the two lowest states, $B(E2; 2_1^+\to 0_1^+)$. Interband transitions 
are listed in Table 1.
\label{f5}}
\end{figure}

\begin{table}[htb]
\centering
\caption{Interband $B(E2; L_i\to L_f)$ transition rates for the X(3) model
\cite{X3}, normalized to the one between the two lowest states, 
$B(E2; 2_1^+\to 0_1^+)$.}
\medskip
\begin{tabular}{c r c r c r c r c r c r}         
\hline\hline\noalign{\smallskip}
$L_i, L_f$ & X(3) & $L_i, L_f$ & X(3) & $L_i, L_f$ & X(3) &
$L_i, L_f$ & X(3) & $L_i, L_f$ & X(3) & $L_i, L_f$ & X(3) \\
\noalign{\smallskip}\hline\noalign{\smallskip}
$ 0_2,  2_1$ &164.0 &  & & & &
$ 0_3,  2_2$ &209.1 &  & & & \\
$ 2_2,  4_1$ & 64.5 & $  2_2,  2_1$ &12.4 & $  2_2,  0_1$ & 0.54 &
$ 2_3,  4_2$ & 92.0 & $  2_3,  2_2$ &16.2 & $  2_3,  0_2$ & 0.67 \\
$ 4_2,  6_1$ & 42.2 & $  4_2,  4_1$ & 8.6 & $  4_2,  2_1$ & 0.43 &
$ 4_3,  6_2$ & 65.3 & $  4_3,  4_2$ &12.2 & $  4_3,  2_2$ & 0.47 \\
$ 6_2,  8_1$ & 31.1 & $  6_2,  6_1$ & 6.7 & $  6_2,  4_1$ & 0.51 &
$ 6_3,  8_2$ & 50.9 & $  6_3,  6_2$ &10.1 & $  6_3,  4_2$ & 0.52 \\
$ 8_2, 10_1$ & 24.4 & $  8_2,  8_1$ & 5.5 & $  8_2,  6_1$ & 0.56 &
$ 8_3, 10_2$ & 41.6 & $  8_3,  8_2$ & 8.6 & $  8_3,  6_2$ & 0.57 \\
$10_2, 12_1$ & 19.9 & $ 10_2, 10_1$ & 4.7 & $ 10_2,  8_1$ & 0.59 &
$10_3, 12_2$ & 35.0 & $ 10_3, 10_2$ & 7.5 & $ 10_3,  8_2$ & 0.61 \\
$12_2, 14_1$ & 16.6 & $ 12_2, 12_1$ & 4.0 & $ 12_2, 10_1$ & 0.60 &
$12_3, 14_2$ & 30.1 & $ 12_3, 12_2$ & 6.6 & $ 12_3, 10_2$ & 0.63 \\
$14_2, 16_1$ & 14.2 & $ 14_2, 14_1$ & 3.5 & $ 14_2, 12_1$ & 0.60 &
$14_3, 16_2$ & 26.3 & $ 14_3, 14_2$ & 5.9 & $ 14_3, 12_2$ & 0.65 \\
$16_2, 18_1$ & 12.3 & $ 16_2, 16_1$ & 3.1 & $ 16_2, 14_1$ & 0.60 &
$16_3, 18_2$ & 23.3 & $ 16_3, 16_2$ & 5.4 & $ 16_3, 14_2$ & 0.66 \\
$18_2, 20_1$ & 10.9 & $ 18_2, 18_1$ & 2.8 & $ 18_2, 16_1$ & 0.59 &
$18_3, 20_2$ & 20.8 & $ 18_3, 18_2$ & 4.9 & $ 18_3, 16_2$ & 0.66 \\
$20_2, 22_1$ &  9.7 & $ 20_2, 20_1$ & 2.5 & $ 20_2, 18_1$ & 0.58 &
$20_3, 22_2$ & 18.8 & $ 20_3, 20_2$ & 4.5 & $ 20_3, 18_2$ & 0.66 \\
\noalign{\smallskip}\hline\hline
\end{tabular}
\end{table}

\begin{figure}[ht]   
\rotatebox{270}{\includegraphics[width=55mm]{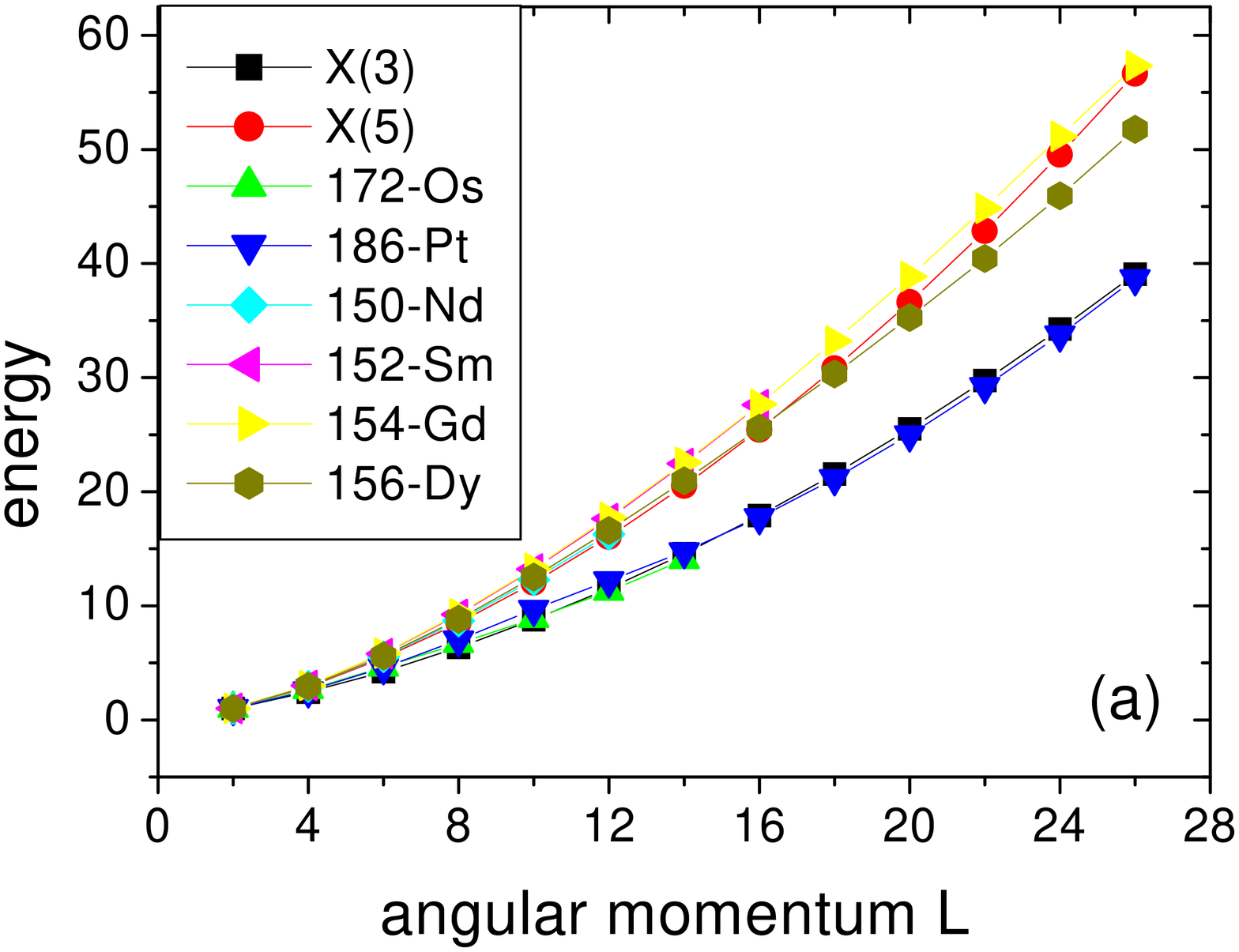}} 
\rotatebox{270}{\includegraphics[width=55mm]{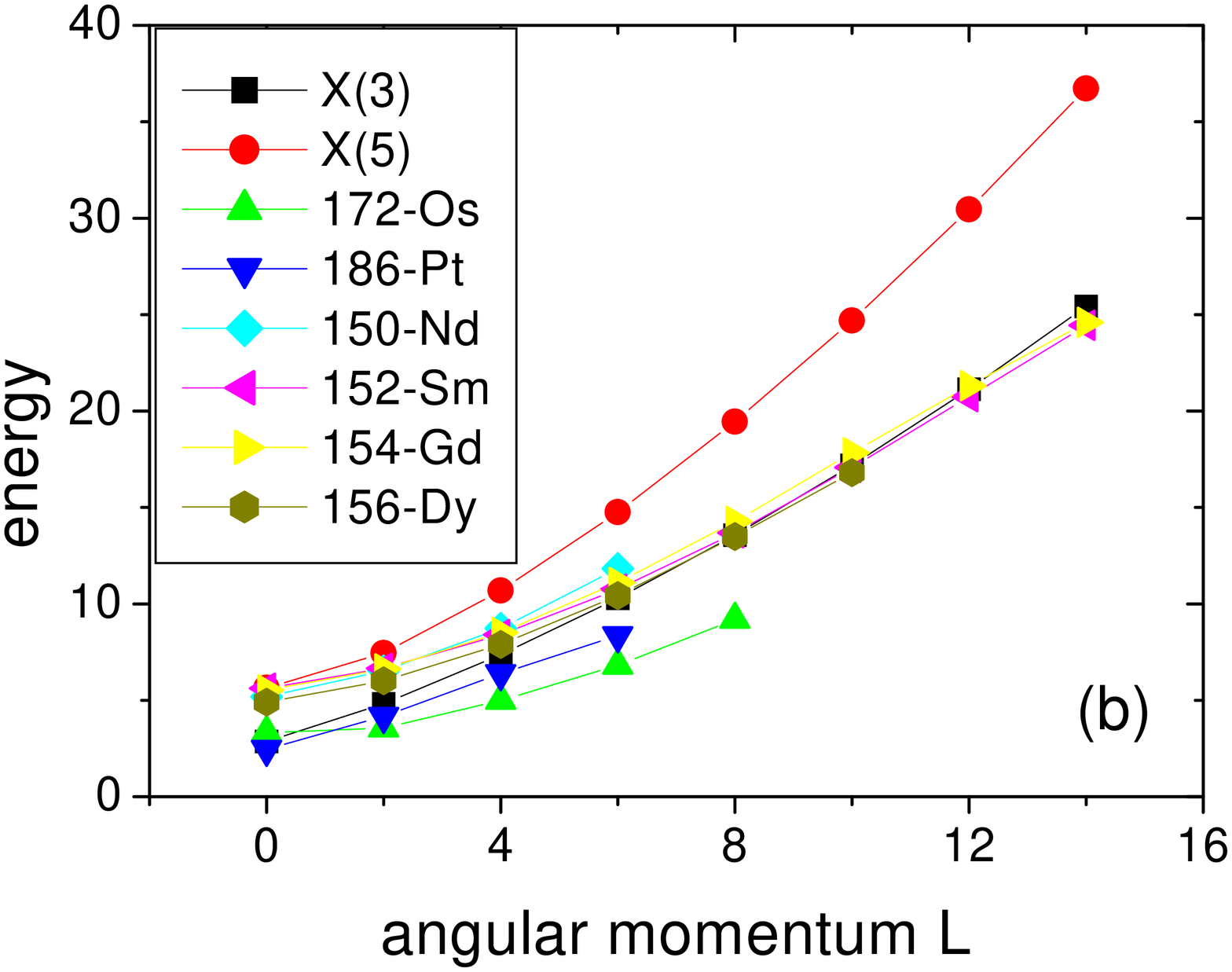}} 
\caption{
(a) Energy levels of the ground state bands of the X(3) \cite{X3} and 
X(5) \cite{IacX5} 
models, compared to experimental data for $^{172}$Os \cite{172Os}, 
$^{186}$Pt \cite{186Pt}, $^{150}$Nd \cite{150Nd}, $^{152}$Sm \cite{152Sm}, 
$^{154}$Gd \cite{154Gd}, and $^{156}$Dy \cite{156Dy}. The levels of each 
band are normalized to the $2_1^+$ state. 
(b) Same for the $\beta_1$-bands, also normalized to the $2_1^+$ state. 
\label{f6}}
\end{figure}  

It should be remarked that in all of the above mentioned cases the Bessel
eigenfunctions obtained are of the form $J_\nu(k \beta)$, with $\nu$ being 
of the form 
\begin{equation}
\nu=\sqrt{\Lambda + \left({n-2\over 2}\right)^2 },
\end{equation}
where $n$ is the number of dimensions entering in the problem, while 
$\Lambda = L(L+1)/3$ in the cases of X(3) \cite{X3} and X(5) \cite{IacX5}, 
$\Lambda = [L(L+4)+3 n_w (2L-n_w)]/4$ in the cases of Z(4) \cite{Z4} and 
Z(5) \cite{Z5}, 
with $n_w =L-\alpha$ being the wobbling quantum number \cite{BM}, and 
$\Lambda = \tau(\tau+3)$ in the case of E(5), with $\tau$ being the seniority
quantum number characterizing the irreducible representations of the SO(5) 
subalgebra of E(5) \cite{IacE5}. In the corresponding ground state bands 
one has $n_w=0$ and $\tau=L/2$. 

One should also notice that in all of the above cases the spectrum is 
determined by the boundary condition that the eigenfunctions have to vanish 
at the boundaries of the infinite square well potential. As a result, 
in addition to the other relevant quantum numbers, the energy levels are 
characterized by $s$, the order of the corresponding root of the relevant 
Bessel function.

It should also be mentioned that all the $\beta$-equations mentioned above 
are also soluble \cite{Elliott,Rowe} if the infinite square well potential is 
substituted by a Davidson 
potential \cite{Dav} of the form $u(\beta) =\beta^2 +\beta_0^4/\beta^2$, where 
$\beta_0$ is the minimum of the potential, the eigenfunctions being Laguerre 
polynomials instead of Bessel functions in this case. A variational procedure 
has been developed \cite{varPLB,varPRC}, in which the first derivative of 
various collective quantities
is maximized with respect to the parameter $\beta_0$, leading to the 
E(5), X(5), Z(5), and Z(4) results in the corresponding cases, an example 
being shown in Fig. 7(a). The solutions corresponding to the Davidson 
potentials lead to monoparametric curves connecting various collective 
quantities, an example being shown in Fig. 7(b), where agreement with 
experimental data is very good. 

\begin{figure}[htb]  
\rotatebox{270}{\includegraphics[width=55mm]{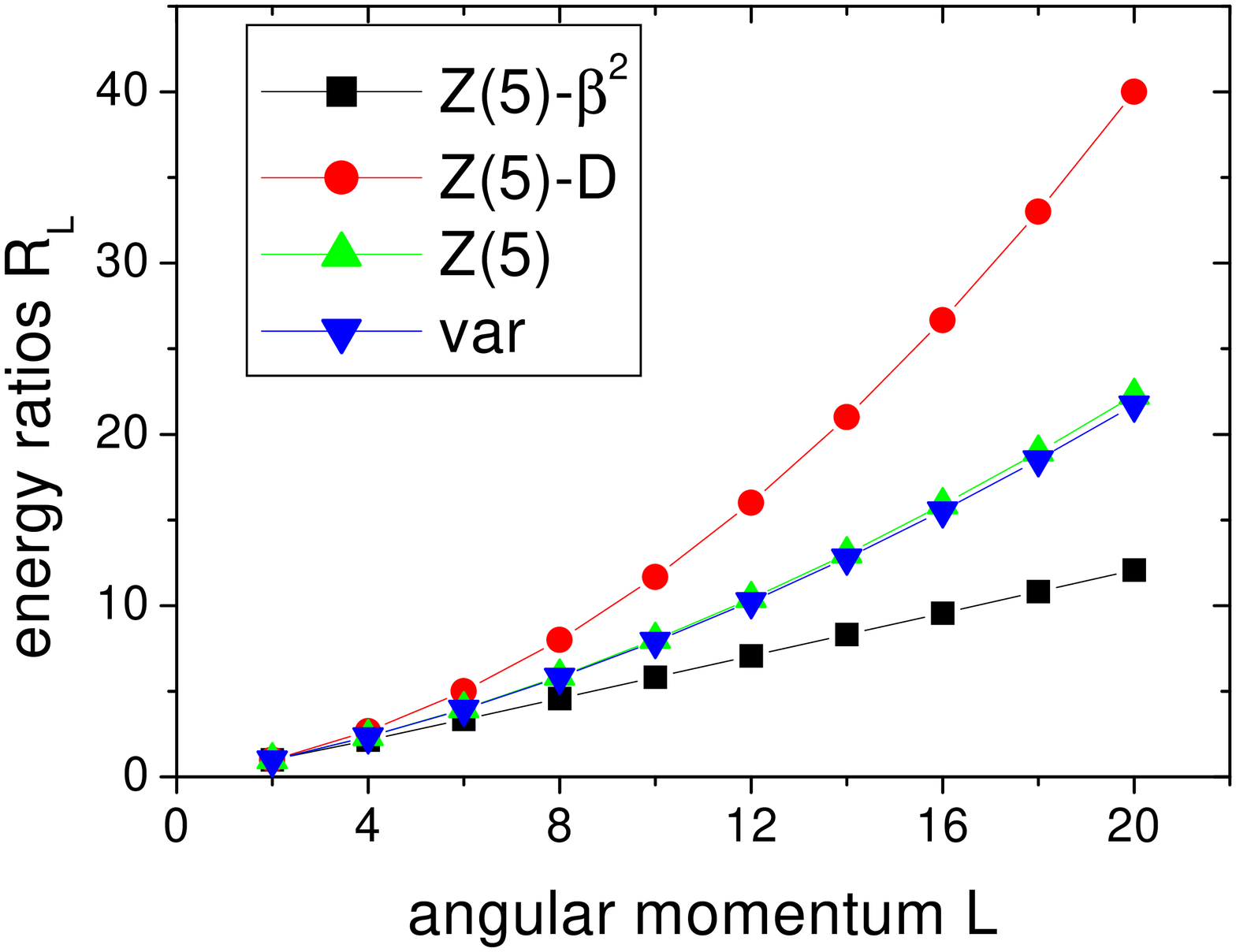}}
\rotatebox{270}{\includegraphics[width=55mm]{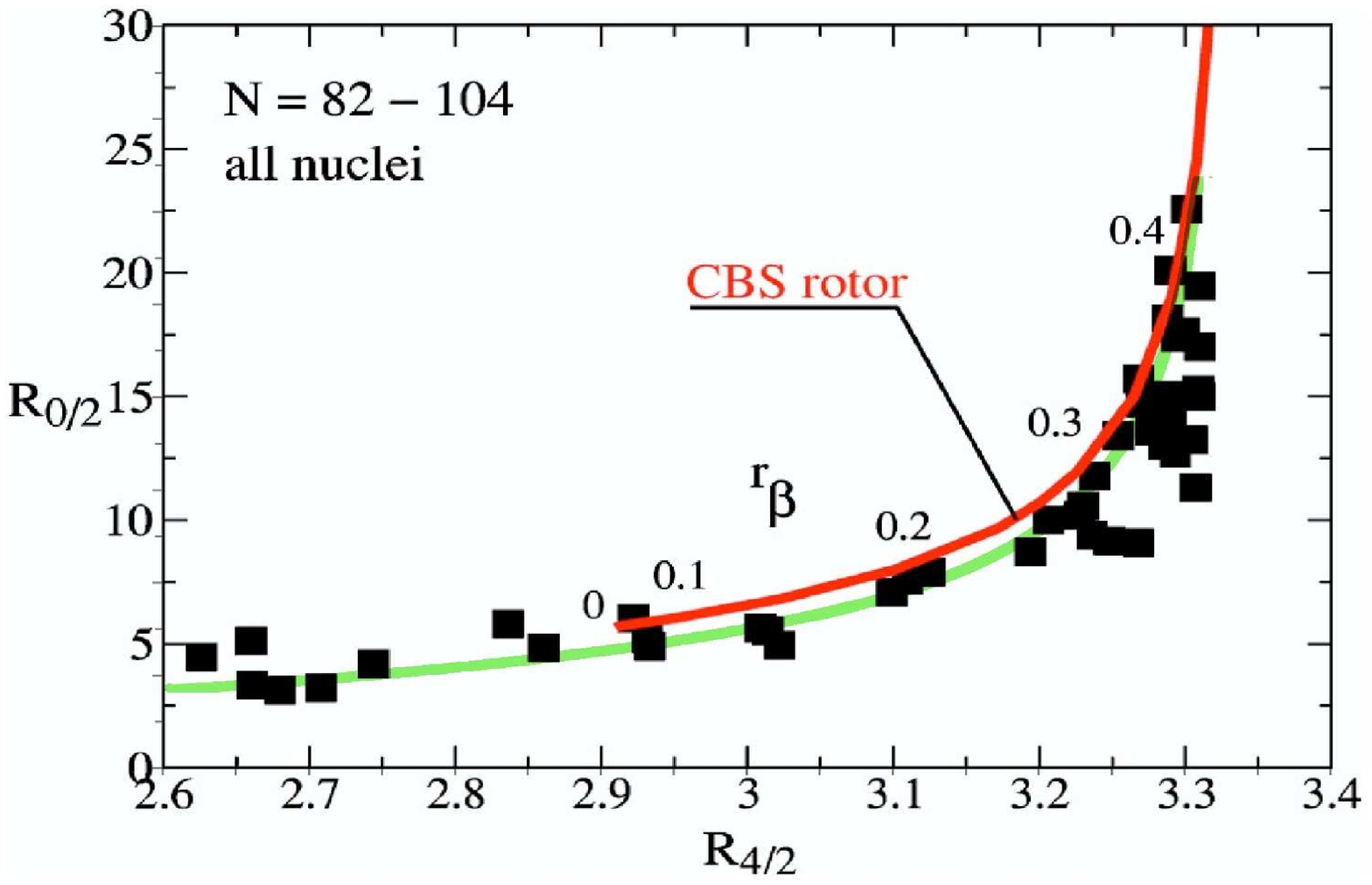}}
\caption{(a) Values of the ratio $R_L=E(L)/E(2)$ for the ground state band
obtained through the variational procedure (labeled by ``var'') using 
Davidson potentials in the Z(5) framework, compared to the values provided 
by the Z(5)-$\beta^2$, Z(5)-D$(\beta_0)$ with $\beta_0\to \infty$ (labeled 
as Z(5)-D), and Z(5) models (see \cite{X5} for the relevant terminology). 
(b) Monoparametric curves of $R_{0/2}= E(0_2^+)/E(2_1^+)$ versus 
$R_{4/2}=E(4_1^+)/E(2_1^+)$ resulting from the ``confined $\beta$-soft''
(CBS) rotor \cite{PG} (labeled as CBS rotor, with the values of the parameter
$r_\beta$ given along the curve) and from Davidson potentials in the X(5) 
framework \cite{varPLB,varPRC}, compared to experimental data (taken from Ref. 
\cite{PG}). 
\label{f7}}
\end{figure}

Concerning future work, the clarification of the algebraic structure of the 
exactly soluble models X(3) and Z(4), as a prelude for the understanding 
of the algebraic structure of the approximate solutions X(5) and Z(5), 
is a challenging problem. The construction of analytical models including 
the octupole degree of freedom \cite{AQOA} and/or the dipole degree of freedom 
is also receiving attention. 

One of the authors (IY) is thankful to the Turkish Atomic Energy Authority 
(TAEK) for support under project number 04K120100-4.

\end{document}